\documentclass{ws-procs9x6}

\def\MeV{{\rm Me\!V}}
\def\GeV{{\rm Ge\!V}}
\def\msbar{{\overline{\rm MS}}}

\begin{document}

\title{LOW ENERGY CONSTANTS FROM THE MILC COLLABORATION\footnote{Talk presented
by C.\ Bernard at  Chiral Dynamics 2006, Duke University, Sept. 18-22.}}

\author{C. Bernard}

\address{Department of Physics,  Washington University,\\
Saint Louis, Missouri 63130, USA\\
E-mail: cb@wustl.edu}

\author{C.\ DeTar and L.\ Levkova}
        
\address{Department of Physics,  University of Utah\\ 
	Salt Lake City, Utah 84112, USA}

\author{Steven Gottlieb}
        
\address{Department of Physics,  Indiana University\\
 Bloomington, Indiana 46405, USA}

\author{U.M.\ Heller}
 
\address{American Physical Society\\
 One Research Road, Ridge, New York 11961, USA}

\author{J.E.\ Hetrick}

\address{Department of Physics,  University of the Pacific\\
 Stockton, California 95211, USA}

\author{J. Osborn}

\address{Department of Physics,  Boston University\\
 Boston, Massachusetts 02215, USA}

\author{D.\ Renner and D.\ Toussaint}

\address{Department of Physics, University of Arizona\\
 Tucson, Arizona 85721, USA}

\author{R.\ Sugar}

\address{Department of Physics, University of California\\
 Santa Barbara, California 93106, USA}

\begin{abstract}
We present preliminary updates of results for QCD low energy constants.
\end{abstract}

\keywords{chiral dynamics, lattice QCD}

\bodymatter

\section*{}\label{aba:sec1}
The MILC Collaboration is in the process of updating its 
simulations \cite{Aubin:2004fs}
of QCD using three dynamical light flavors ($u,d,s$) 
of improved staggered quarks.
New lattice data \cite{Bernard:2006wx}
includes a complete ensemble with average $u,d$ mass
$\hat m = 0.1 m_s$ and lattice spacing $a\approx 0.09\;$fm, and a partial
ensemble at $\hat m = 0.4  m_s$ and  $a\approx 0.06\;$fm, as well
as several new coarser ensembles at $a\approx0.15\;$fm, with a
range of sea quark masses.

Some of our preliminary results are:
\begin{eqnarray*}
f_\pi  =  128.6 \pm 0.4\pm 3.0 \; \MeV; &&  
f_K  =    155.3 \pm 0.4\pm 3.1 \; \MeV; \\
f_K/f_\pi   = 1.208(2)({}^{+\phantom{1}7}_{-14}) &\Rightarrow& 
|V_{us}|=0.2223({}^{+26}_{-14}); \\
   f_\pi/f_{2} =  1.050(3)(10); && f_{2}/f_{3}  =  1.09(4)(4); \\
2L_6 - L_4 = 0.5(1)(2); &&  
2L_8 - L_5 = -0.1(1)(1); \\
L_4 = 0.1(2)(3); && L_5 =2.0(3)(2);  \\
 \langle\bar uu\rangle_{2} =   -(\, 276(2)(7)(5)\;\MeV\,)^3 ;&&
\langle\bar uu\rangle_{3}  =  -(\, 247(10)(15)(4)\;\MeV\,)^3 ; \\
 \langle\bar uu\rangle_{2}/\langle\bar uu\rangle_{3}  =  1.38(15)(22); && \\
m_s^\msbar = 90(0)(5)(4)(0)\;\MeV; &&  \hat m^\msbar = 3.3(0)(2)(2)(0)\;\MeV; \\
m_u^\msbar = 2.0(0)(1)(2)(1)\;\MeV; &&  m_d^\msbar = 4.6(0)(2)(2)(1)\;\MeV; \\
m_s/\hat m = 27.2(0)(4)(0)(0); &&  m_u/m_d = 0.42(0)(1)(0)(4)\;.
\end{eqnarray*}
The errors are statistical, systematic (from the simulations), and, where relevant, perturbative (from two-loop perturbation theory \cite{Mason:2005bj})
and electromagnetic. 
The quantity $f_2$ ($f_3$) represents the three-flavor decay constant
in the two (three) flavor chiral limit, and 
$\langle\bar uu\rangle_{2}$ ($\langle\bar uu\rangle_{3}$)
is the corresponding condensate.
The low energy constants $L_i$ are evaluated at chiral scale  
$m_\eta$, and the condensates and masses are in the
$\msbar$ scheme at scale $2\,\GeV$.

Our computations use the ``fourth root trick'' to reduce the staggered
degrees of freedom.  While this gives an action that is non-local
at non-zero lattice spacing \cite{Bernard:2006ee}, recent work
\cite{Shamir:2006nj,Bernard:2006zw,Bernard:2006qt} indicates that
the desired local QCD theory is obtained in the continuum limit.  In addition,
the complicated effects of the non-locality at non-zero lattice spacing 
can be controlled and extrapolated away with ``rooted staggered chiral perturbation
theory.'' \cite{Bernard:2006zw,Bernard:2006qt}  More work is still required to check
the assumptions and arguments in 
Refs.~\refcite{Shamir:2006nj,Bernard:2006zw,Bernard:2006qt}; for a comprehensive
review of the status and prospects, 
see Ref.~\refcite{Sharpe:2006re}.

\bibliographystyle{ws-procs9x6}
\bibliography{MILC_LEC}

\end{document}